\documentclass[letterpaper]{article}
\usepackage[T1]{fontenc}
\usepackage{spconf,amsmath,amssymb,graphicx,booktabs,balance}
\usepackage[hidelinks]{hyperref}

\newcommand{\dB}{\,dB}
\newcommand{\vfig}[1]{figs/#1}
\title{Quieter Than the Room:\\Representation Drift and Task Robustness in Speech Encoders}
\twoauthors
  {Vsevolod Kovalev}
  {Boston University, Boston, MA, USA\\
   vsevolod.kovalev@outlook.com}
  {Pranay Manocha}
  {Princeton University, Princeton, NJ, USA\\
   pranay@symbal.ai}

\makeatletter
\def\ps@arxivnotice{
  \let\@oddhead\@empty
  \let\@evenhead\@empty
  \def\@oddfoot{\parbox{\textwidth}{\footnotesize\normalfont
    This work has been submitted to the IEEE for possible publication. Copyright may be transferred without notice, after which this version may no longer be accessible.}}
  \let\@evenfoot\@oddfoot
}
\makeatother

\begin{document}
\ninept
\maketitle
\thispagestyle{arxivnotice}

\begin{abstract}
Non-speech interference can change a speech representation without causing
comparable task loss. We test eight frozen encoders on four tasks, adding
non-speech sounds throughout recordings, during speech, or in pauses.
Under whole-recording interference, embedding drift tracks task loss across
seven sounds, with mean Spearman correlations of 0.81--0.88.
Moving the same sound between speech and pauses changes this pattern.
At quiet to moderate levels, pause interference produces larger drift,
while speech interference usually causes greater loss on intent recognition,
speaker verification and speech recognition. Emotion recognition shows a
weaker placement effect. Pause interference also changes speech-frame
representations beyond the injected region. Even below the estimated
recording background, interference can change embeddings as much as repeated
speech takes do. Drift helps rank the effects of different
sounds, but larger drift does not consistently indicate greater task loss.
\end{abstract}


\section{Introduction}
\label{sec:intro}

Speech encoders support recognition, speaker and paralinguistic tasks
\cite{yang2021superb,turian2022hear}. Their inputs often include non-speech
sound in homes and distant-microphone settings
\cite{bastianelli2020slurp,watanabe2020chime6,nandwana2020voices}.
We call this added non-speech sound \emph{interference} and study how it
changes both encoder representations and task performance.
Robustness evaluations commonly report downstream scores
\cite{shah2024srb,yang2024largescale,arcos2026dimensionality}, which can remain
stable even when the underlying representation changes substantially.
Speech Robust Bench evaluates ASR under environmental noise
\cite{shah2024srb}; distorted SUPERB uses 3, 6 or 9\dB{}
\cite{yang2024largescale}. We extend the comparison to quieter interference,
including levels below a recording's estimated background.

The experiments vary both sound type and placement. Time-averaged embeddings combine
speech and non-speech frames, and contextual encoders carry information
across time \cite{meng2025context}. Non-speech audio can affect ASR
\cite{baranski2025hallucinations}; non-speech augmentation has also improved
transfer to interview audio for scripted-versus-spontaneous speech detection
\cite{kovalev2026seam}. We first compare sounds added throughout recordings,
then move the same sounds between speech and pauses to test whether larger
embedding changes accompany larger drops in task performance.

Controlled speech variation provides a reference for noise-induced drift.
Speech quality metrics based on representations
\cite{ogg2026s3qa,saeki2024speechbertscore} and distances learned from
perceptual judgments \cite{manocha2020jnd} motivate this use of a reference.
Repeated takes and delivery changes in RAVDESS, with speaker and words
fixed, provide an encoder-specific scale. Task loss is measured separately:
an embedding can change in ways that have little effect on a task decision.

We find that drift tracks task loss across sounds under whole-recording
interference, but often gives the opposite ordering when placement changes.
Quiet pause interference can produce drift comparable to repeated takes,
and its effects extend into speech-frame representations.

\section{Evaluation}
\label{sec:method}
\subsection{Encoders and representation drift}

We evaluate eight frozen encoders: HuBERT Base and Large, WavLM Base,
Base+ and Large, wav2vec~2.0 Large Robust, Whisper Medium and Whisper
Large-v2
\cite{hsu2021hubert,chen2022wavlm,baevski2020wav2vec2,hsu2021robust,radford2023whisper}.
Audio is processed at its original length with batch size one; perturbations
do not extend the recordings. For layer $\ell$, the average over $T$
frames is $e_\ell=T^{-1}\sum_t h_\ell[t]$.
With $e,e'$ denoting clean and perturbed embeddings, drift is their cosine
distance:
\begin{equation}
d(e,e')=1-\frac{\langle e,e'\rangle}{\lVert e\rVert_2\lVert e'\rVert_2}.
\label{eq:drift}
\end{equation}
We report the last layer.
SpeechBERTScore, a frame-matching metric, agrees with cosine on the
reference speech pairs (Spearman $\rho=0.86$--$0.92$). We also measure
drift after each task head's learned layer sum.

Last-layer distances for wav2vec~2.0 Large Robust are about
$10^3$ smaller than those of the other encoders, making reference crossings
numerically unstable in our pipeline. We include it in task evaluation,
but exclude it from drift summaries and the paired placement comparison.

\subsection{Speech variation as a reference}

RAVDESS contains scripted speech from 24 actors, with repeated takes and
controlled changes in emotion and intensity~\cite{livingstone2018ravdess}.
We compare recordings of the same speaker saying the same sentence.
For each encoder and layer, the repeat-take reference $L$ is the
smallest mean distance between repeated takes within a delivery condition.
The delivery-variation reference $U$ is the largest mean distance
between delivery conditions. Neutral takes give the smallest repeat distance;
the pair that defines $U$ always includes strongly fearful speech.

The interval from $L$ to $U$ forms a corridor. After division by $L$,
its floor is 1 and its ceiling is $U/L$ (6.5--8.2 across encoders).
A crossing SNR is the level at which drift reaches either reference as
interference becomes louder. Last-layer and layer-averaged ceiling crossings
differ by a median 6\dB, so we report last-layer cosine as the main distance
metric. The corridor floor and ceiling are reference cosine distances.
They set the scale for normalization and crossing SNRs; we compute drift and task scores without these anchors.

We assess sensitivity to actor selection with 1,000 random 12/12-actor
splits and 2,000 actor bootstraps, reselecting reference conditions and
reweighting drift curves in each resample. We rank fourteen recorded sounds
and white noise under whole-recording interference on calm-normal clips,
using the seven-encoder median ceiling-crossing SNR.
Split-half Spearman agreement has median 0.975 and minimum 0.906; the
male/female split gives 0.968. For sounds crossing in the full sample,
bootstrap crossing SD averages 1.99\dB{} over draws that cross.

\begin{figure}[t]
\centering
\includegraphics[width=\linewidth]{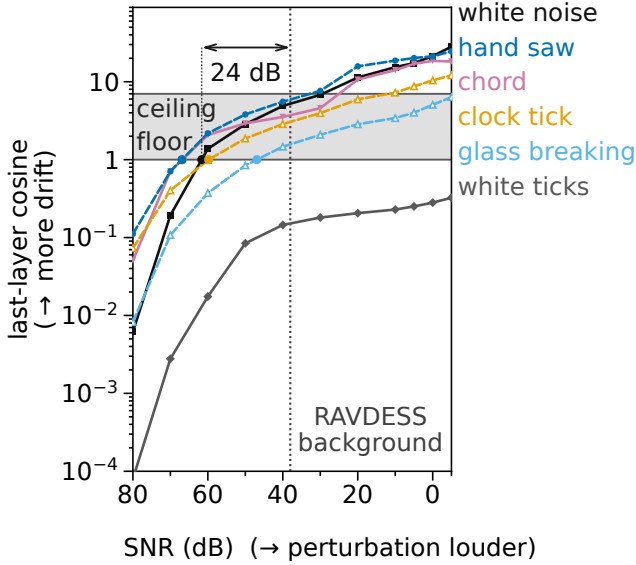}
\caption{Drift under whole-recording interference on neutral RAVDESS,
normalized by each encoder's repeat-take reference (seven-encoder median).
The shaded corridor spans repeat-take and delivery variation.
Solid/dashed curves denote generated/recorded sounds; filled markers mark
floor crossings. The dotted line is the estimated background.}
\label{fig:corridor}
\end{figure}

\subsection{Sounds, levels and placement}

We compare nine generated sounds with fourteen recorded ESC-50 classes
\cite{piczak2015esc}, and combine white, pink and tonal signals with continuous,
gated and tick-like timing, at duty cycles $\delta\in\{1,0.10,0.006\}$.
The recorded set includes nine classes from a $3\times3$ grid of spectral flatness
and waveform kurtosis, plus five classes for comparisons at similar duty.
Flatness describes how evenly spectral energy is distributed; kurtosis
describes waveform impulsiveness. Recorded duty is the fraction of 10\,ms
frames within 30\dB{} of peak energy. Selected class medians range from
0.23 to 1.00. Since no ESC-50 class has median duty below 0.13, we use generated
sounds to test sparser interference.

The SNR sweep covers eleven levels from 80 to $-5$\dB. Let $s$ be the original
waveform, $p$ the added sound, $\mathcal V$ the speech samples marked by
Silero VAD \cite{silero2024vad}, and $\mathcal G$ the samples where $p$
is active. Using root-mean-square amplitude (RMS) over these samples,
\begin{equation}
\mathrm{SNR}_{\mathrm{tag}}=20\log_{10}
\frac{\mathrm{RMS}(s,\mathcal V)}{\mathrm{RMS}(p,\mathcal G)}.
\label{eq:snr}
\end{equation}
Higher SNR means quieter interference. We add sounds throughout the
recording, only in speech ($\mathcal G\subseteq\mathcal V$), or only in
pauses ($\mathcal G\cap\mathcal V=\emptyset$). Pauses are VAD-defined
non-speech regions and may contain other sounds. Active RMS matches the
level while each sound is present; sparse sounds inject less total energy.
On RAVDESS, speech and pauses occupy 0.47 and 0.53 of a clip, respectively,
so the two placements inject nearly equal energy.

Whole-clip RMS includes inactive samples, making sparse sounds louder at
the same nominal SNR. For generated gated sounds, the two conventions
differ by 10--22\dB. With whole-clip RMS, kurtosis-matched generated and
recorded sounds still cross the floor at least 12\dB{} apart.

For clip $i$, we estimate background level $b_i$ from the median RMS of
20\,ms non-speech frames, in dB relative to speech RMS. RAVDESS's corpus
estimate is $-37.9$\dB, equivalent to a tagged SNR of 37.9\dB.
To account for the quieter background in pauses, we also express each
injection relative to the region it occupies:
\begin{equation}
\mathrm{SNR}_{\mathrm{rel},i}=\begin{cases}
\mathrm{SNR}_{\mathrm{tag}}, & \text{speech placement},\\
\mathrm{SNR}_{\mathrm{tag}}+b_i, & \text{pause placement}.
\end{cases}
\label{eq:relative}
\end{equation}
For $b_i=-38$\dB, pause interference at 60\dB{} tagged SNR is 22\dB{}
below background. This conversion applies to existing observations per clip;
elsewhere, SNR refers to Eq.~\ref{eq:snr}.

We use neutral RAVDESS clips for placement and Fig.~\ref{fig:corridor},
and calm deliveries for generated--recorded comparisons. Encoders receive
identical perturbed audio; four perturbation seeds give between-draw
variation of 5--6\% of drift.

To measure changes beyond the injection, we pool only frames whose local
input spans contain no injected samples. Per-frame cosine drift is also
measured against distance from the nearest injection.

\subsection{Downstream evaluation}

We select four SUPERB tasks covering semantics, paralinguistics, speaker
identity and content: intent classification (IC) on Fluent Speech Commands
(FSC), emotion recognition (ER) on IEMOCAP, speaker verification (ASV) on
VoxCeleb1, and automatic speech recognition (ASR) on LibriSpeech
\cite{lugosch2019fluent,busso2008iemocap,nagrani2017voxceleb,panayotov2015librispeech}.
Encoders remain frozen, and task heads are trained on clean audio.
Clean scores differ from published SUPERB results by at most 0.3 accuracy
points for IC, 1.6 for ER, about 1 equal-error-rate (EER) point for ASV,
and 1.1--2.1 word-error-rate (WER) points for ASR. ASV and ASR heads use
fixed training budgets and remain undertrained by our design. Changes from
each head's own clean score quantify its response to interference. Task loss is an accuracy decrease for IC/ER or an EER/WER
increase for ASV/ASR, in percentage points.

The downstream tests cover white noise and six recorded sounds under all three placements:
church bells, hand saw, sheep, insects, clock tick and glass breaking.
The first four recordings are dense ($\delta>0.6$); the last two are sparse.
For speech--pause comparisons, we select clips with similar pause-to-speech
duration ratios across corpora (0.44--0.52).
Figure~\ref{fig:placement} uses the same seven encoders for drift and task loss.

\begin{figure*}[t]
\centering
\includegraphics[width=\textwidth]{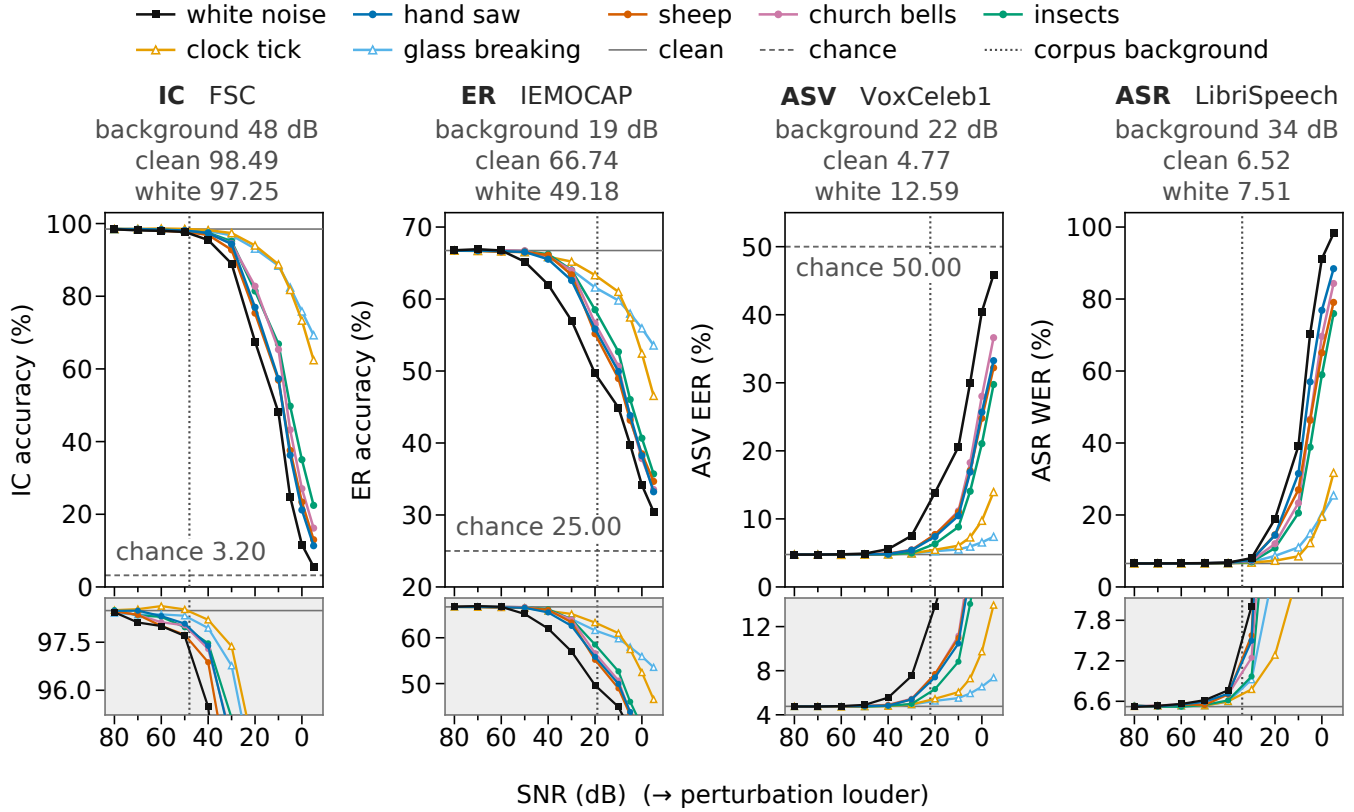}
\caption{Task performance under whole-recording interference from seven
sounds (eight-encoder means). Dense recordings generally cause greater
losses than sparse recordings as interference grows louder. Dotted lines
mark estimated corpus backgrounds; horizontal solid/dashed lines denote
clean/chance performance. Lower panels magnify the region near clean
performance. Squares denote white noise, circles dense recordings and
triangles sparse recordings. Above each panel, the readouts give the
background SNR, clean score and white-noise score at that SNR.}
\label{fig:uniform}
\end{figure*}

\section{Results}
\label{sec:results}

\subsection{Quiet interference can match speech variation}
\label{ssec:reference}

White noise, pink noise and a chord added continuously in pauses reach
the repeat-take reference 24--34\dB{} below the estimated RAVDESS
background on the five self-supervised encoders. Whisper crossings are closer to the background,
at 4--18\dB{} below it. Replacing the neutral repeat-take reference with
the calm reference leaves these margins almost unchanged.

Temporal coverage strongly affects drift under whole-recording interference
(Fig.~\ref{fig:corridor}). All fourteen recorded classes cross the floor on
all seven encoders; the sparsest generated signals usually remain below it.
Across 20 sounds that cross, duty correlates with crossing SNR at
$\rho=0.85$: sounds active for longer reach the reference at quieter levels.
Generated and recorded sounds matched on duty cross in similar SNR ranges;
matching on kurtosis leaves larger gaps. Greater duty also means more
injected energy at a fixed active SNR, so these effects are coupled.

Pause interference has greater contrast against its local background.
Using region-relative SNR (Eq.~\ref{eq:relative}) reduces the speech--pause
gap. The most tonal sounds still reach the floor at levels 8--12\dB{}
quieter in pauses than in speech. Both Whisper encoders, however, are more
sensitive to speech interference on this scale.

\begin{figure*}[t]
\centering
\includegraphics[width=\textwidth]{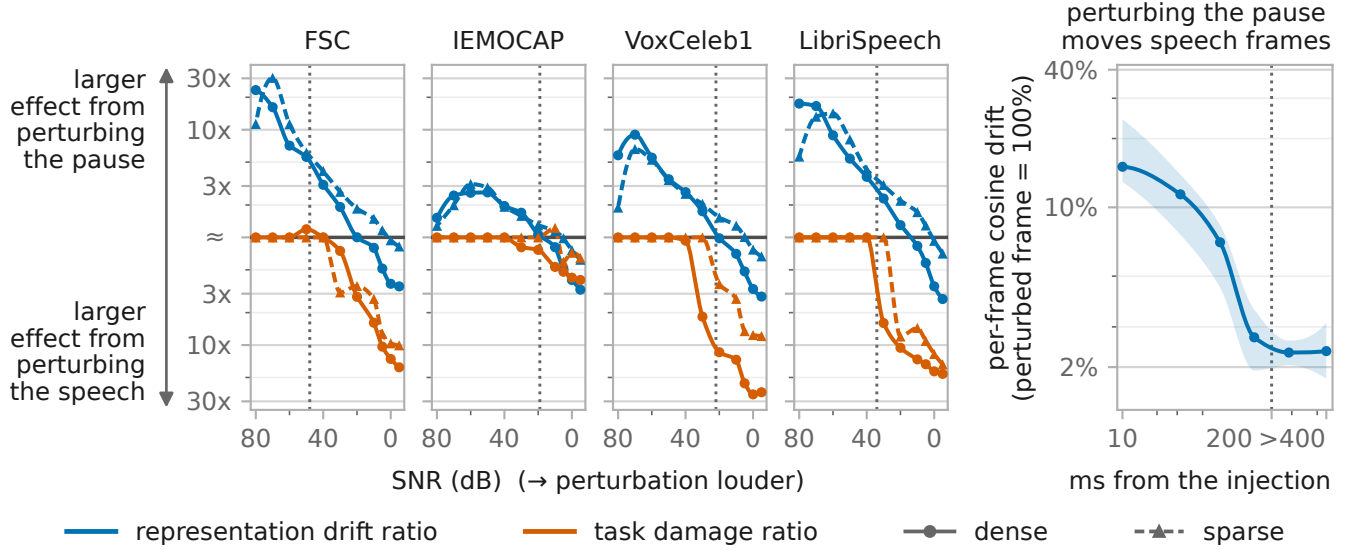}
\caption{Left: seven sounds placed in speech or pauses, with similar pause
proportions across corpora. The mirrored scale shows how many times larger
the effect is in pauses (above equality) or speech (below). Drift uses
a seven-encoder median and task loss a seven-encoder mean, pooled within
sound groups. The equality line also serves as a plotting baseline where
task losses are too small for a stable ratio. Dotted lines
mark corpus backgrounds. Right: a separate RAVDESS experiment shows
speech-frame drift beyond pause injections, relative to drift in perturbed
frames (six dense sounds at 60\dB{} SNR; median and range).}
\label{fig:placement}
\end{figure*}

\subsection{Drift tracks damage across sounds}
\label{ssec:consequence}

Under whole-recording interference, dense sounds generally cause larger
task losses than sparse sounds (Fig.~\ref{fig:uniform}). Effects near the
estimated background vary by task. At IEMOCAP's background, the most
damaging dense recording reduces emotion accuracy by 9--13 points on
every encoder. Changes on the other tasks are smaller near their respective
backgrounds. As interference grows louder, the separation between dense
and sparse recordings becomes more clear across all four tasks.

\begin{table}[t]
\centering
\caption{Mean Spearman correlation with task loss across four tasks.
Each correlation ranks seven sounds under whole-recording interference
at the stated SNR. RAVDESS crossings are reused across tasks and levels.}
\label{tab:predictors}
\begin{tabular*}{\linewidth}{@{\extracolsep{\fill}}lrrr@{}}
\toprule
Measure & 20\dB & 10\dB & 0\dB \\
\midrule
Duty & 0.70 & 0.66 & 0.65 \\
Drift, task corpus & 0.81 & 0.86 & 0.88 \\
Drift, RAVDESS & 0.77 & 0.77 & 0.84 \\
Repeat-take crossing & 0.47 & 0.53 & 0.60 \\
Delivery-variation crossing & 0.82 & 0.88 & 0.88 \\
\bottomrule
\end{tabular*}
\end{table}

Representation drift captures more of this ordering than duty alone
(Table~\ref{tab:predictors}). Drift measured on each task corpus reaches
mean Spearman correlations of 0.81--0.88 with task loss, compared with
0.65--0.70 for duty. Delivery-reference (ceiling) crossings measured on RAVDESS
perform similarly on these seven sounds, suggesting that some of the ordering transfers
across corpora. Repeat-take (floor) crossings correlate less well.

\subsection{Placement separates drift from task damage}
\label{ssec:placement}

We next move the same sounds between speech and pauses
(Fig.~\ref{fig:placement}). At 30\dB{} SNR, pause placement produces larger
drift in nearly every sound--encoder combination on each corpus. IC, ASV
and ASR usually suffer greater losses when the sound overlaps speech.
For example, at 20\dB{} dense sounds increase ASR WER by 5.6 points during
speech, compared with 0.5 in pauses: about 11 times as much loss during
speech (Fig.~\ref{fig:placement}). Here, pause placement causes larger drift
but smaller task loss.

At 40\dB{} SNR and quieter, average losses remain below one point on each
task under both placements, while drift already distinguishes them.
As interference becomes louder, the drift gap narrows; at some levels,
speech placement produces both larger drift and greater task loss.
The mismatch therefore depends on level as well as placement.

Emotion recognition, however, shows a weaker placement effect. On IEMOCAP, speech
is more damaging in only 28 of 49 sound--encoder combinations at 20\dB{},
and the pause-to-speech drift ratio is smaller than on the other corpora.

We also measure drift at the task head's input, after its learned layer
sum. In the white-noise FSC experiment at 20\dB{} SNR, the pause-to-speech
drift ratio falls from 2.4 to 1.7, while speech remains more damaging.
Layer weighting reduces the drift gap but preserves its direction.

A task head may respond weakly to an embedding change even when cosine
drift is large. For a linear score $g(e)=w^\top e+b$, the weights $w$
determine which directions in the embedding affect the score. An embedding
change $e'-e$ changes the score by $w^\top(e'-e)$. If the change is
orthogonal to $w$, the score remains unchanged. Cosine drift compares
$e$ and $e'$ without considering the task weights, so its magnitude alone
cannot determine the score change. Measuring how embedding changes align
with these weights could help explain the observed mismatch. We leave
this test for future work.

\subsection{Pause effects extend into speech frames}
\label{ssec:structure}

Pause interference changes speech-frame representations even when
their local input spans contain no injected samples. We compare
cosine drift after averaging only these speech-frame
representations with drift after averaging all frames. The median
ratio of speech-only drift to all-frame drift is 0.28. The effect
of pause interference therefore persists in the pooled speech
representation, even after the directly perturbed frames are
excluded from the average.

Per-frame drift falls with distance from the injection before leveling
off at about 200\,ms (Fig.~\ref{fig:placement}, right), consistent with
contextual propagation \cite{meng2025context}. Excluding directly perturbed
frames from pooling therefore leaves altered speech representations.

\section{Conclusion}

Embedding drift tracks task loss across the tested sounds under
whole-recording interference. At quiet to moderate levels, pause
interference often produces greater drift, while speech
interference causes larger losses in intent recognition, speaker
verification and ASR. Emotion recognition shows a weaker placement
effect. At louder levels, speech interference can produce both
greater drift and greater task loss. On RAVDESS, interference below
the estimated background can produce drift comparable to repeated
speech takes. Pause interference also alters speech-frame
representations outside the injected region, with changes remaining
when pooling is restricted to speech.

\section{Acknowledgments}
The authors received no funding and declare no conflicts of interest.
Large language models assisted with presentation of author-provided results.
All experiments are the authors' own.

\clearpage
\balance
\section{Compliance with Ethical Standards}
This study is a secondary analysis of existing RAVDESS, ESC-50, FSC,
IEMOCAP, VoxCeleb1 and LibriSpeech data. No new participants were
recruited or human-subject data collected. Institutional ethics review
was not sought.

\bibliographystyle{IEEEbib}
\bibliography{refs,refs_impact}

\begin{thebibliography}{10}

\bibitem{yang2021superb}
{Shu-wen} Yang et~al.,
\newblock ``{SUPERB}: Speech processing universal {PERformance} benchmark,''
\newblock in {\em Proc. Interspeech}, 2021, pp. 1194--1198.

\bibitem{turian2022hear}
J.~Turian et~al.,
\newblock ``{HEAR}: Holistic evaluation of audio representations,''
\newblock in {\em Proc. NeurIPS 2021 Competitions and Demonstrations Track}.
  2022, vol. 176 of {\em Proceedings of Machine Learning Research}, pp.
  125--145, PMLR.

\bibitem{bastianelli2020slurp}
E.~Bastianelli, A.~Vanzo, P.~Swietojanski, and V.~Rieser,
\newblock ``{SLURP}: {A} spoken language understanding resource package,''
\newblock in {\em Proc. Conf. Empirical Methods in Natural Language Processing
  ({EMNLP})}, 2020, pp. 7252--7262.

\bibitem{watanabe2020chime6}
S.~Watanabe et~al.,
\newblock ``{CHiME}-6 challenge: Tackling multispeaker speech recognition for
  unsegmented recordings,''
\newblock in {\em Proc. 6th Int. Workshop on Speech Processing in Everyday
  Environments ({CHiME})}, 2020, pp. 1--7.

\bibitem{nandwana2020voices}
M.~K. Nandwana et~al.,
\newblock ``The {VOiCES} from a distance challenge 2019: Analysis of speaker
  verification results and remaining challenges,''
\newblock in {\em Proc. Speaker and Language Recognition Workshop ({Odyssey})},
  2020, pp. 165--170.

\bibitem{shah2024srb}
M.~A. Shah, D.~Solans Noguero, M.~A. Heikkil{\"a}, B.~Raj, and N.~Kourtellis,
\newblock ``Speech robust bench: A robustness benchmark for speech
  recognition,''
\newblock in {\em Proc. ICLR}, 2025, pp. 27465--27491.

\bibitem{yang2024largescale}
{Shu-wen} Yang et~al.,
\newblock ``A large-scale evaluation of speech foundation models,''
\newblock {\em IEEE/ACM Trans. Audio, Speech, Lang. Process.}, vol. 32, pp.
  2884--2899, 2024.

\bibitem{arcos2026dimensionality}
S.~Arcos-Holzinger, S.~M. Erfani, J.~Bailey, and S.~Khudanpur,
\newblock ``{GRIDS}: Dimensionality-aware anomaly detection in learned
  representations of self-supervised speech models,''
\newblock in {\em Proc. Interspeech [Long Track]}, 2026, pp. 357--366.

\bibitem{meng2025context}
Y.~Meng, S.~Goldwater, and H.~Tang,
\newblock ``Effective context in neural speech models,''
\newblock in {\em Proc. Interspeech}, 2025, pp. 246--250.

\bibitem{baranski2025hallucinations}
M.~Bara{\'n}ski, J.~Jasi{\'n}ski, J.~Bartolewska, S.~Kacprzak, M.~Witkowski,
  and K.~Kowalczyk,
\newblock ``Investigation of {Whisper} {ASR} hallucinations induced by
  non-speech audio,''
\newblock in {\em Proc. IEEE ICASSP}, 2025, pp. 1--5.

\bibitem{kovalev2026seam}
V.~Kovalev and P.~Manocha,
\newblock ``{SEAM}: Shortcut-aware real-time detection of scripted vs.
  spontaneous speech for interview guardrails,''
\newblock in {\em Proc. Interspeech}, 2026, pp. 1391--1395.

\bibitem{ogg2026s3qa}
M.~Ogg, C.~A. Bishop, H.~G. Yi, and S.~R. Robinson,
\newblock ``Self-supervised speech quality assessment ({S3QA}): Leveraging
  speech foundation models for a scalable speech quality metric,''
\newblock {\em J. Acoust. Soc. Amer.}, vol. 159, no. 4, pp. 3662--3673, 2026.

\bibitem{saeki2024speechbertscore}
T.~Saeki, S.~Maiti, S.~Takamichi, S.~Watanabe, and H.~Saruwatari,
\newblock ``{SpeechBERTScore}: Reference-aware automatic evaluation of speech
  generation leveraging {NLP} evaluation metrics,''
\newblock in {\em Proc. Interspeech}, 2024, pp. 4943--4947.

\bibitem{manocha2020jnd}
P.~Manocha, A.~Finkelstein, R.~Zhang, N.~J. Bryan, G.~J. Mysore, and Z.~Jin,
\newblock ``A differentiable perceptual audio metric learned from just
  noticeable differences,''
\newblock in {\em Proc. Interspeech}, 2020, pp. 2852--2856.

\bibitem{hsu2021hubert}
{W.-N.} Hsu, B.~Bolte, {Y.-H.}~H. Tsai, K.~Lakhotia, R.~Salakhutdinov, and
  A.~Mohamed,
\newblock ``{HuBERT}: Self-supervised speech representation learning by masked
  prediction of hidden units,''
\newblock {\em IEEE/ACM Trans. Audio, Speech, Lang. Process.}, vol. 29, pp.
  3451--3460, 2021.

\bibitem{chen2022wavlm}
S.~Chen et~al.,
\newblock ``{WavLM}: Large-scale self-supervised pre-training for full stack
  speech processing,''
\newblock {\em IEEE J. Sel. Topics Signal Process.}, vol. 16, no. 6, pp.
  1505--1518, 2022.

\bibitem{baevski2020wav2vec2}
A.~Baevski, Y.~Zhou, A.~Mohamed, and M.~Auli,
\newblock ``wav2vec 2.0: {A} framework for self-supervised learning of speech
  representations,''
\newblock in {\em Advances in Neural Information Processing Systems
  ({NeurIPS})}, 2020, vol.~33, pp. 12449--12460.

\bibitem{hsu2021robust}
{W.-N.} Hsu et~al.,
\newblock ``Robust wav2vec 2.0: Analyzing domain shift in self-supervised
  pre-training,''
\newblock in {\em Proc. Interspeech}, 2021, pp. 721--725.

\bibitem{radford2023whisper}
A.~Radford, J.~W. Kim, T.~Xu, G.~Brockman, C.~Mcleavey, and I.~Sutskever,
\newblock ``Robust speech recognition via large-scale weak supervision,''
\newblock in {\em Proc. Int. Conf. Machine Learning ({ICML})}, 2023, vol. 202
  of {\em Proceedings of Machine Learning Research}, pp. 28492--28518.

\bibitem{livingstone2018ravdess}
S.~R. Livingstone and F.~A. Russo,
\newblock ``The {Ryerson} audio-visual database of emotional speech and song
  ({RAVDESS}): {A} dynamic, multimodal set of facial and vocal expressions in
  {North American English},''
\newblock {\em PLoS ONE}, vol. 13, no. 5, pp. e0196391, 2018.

\bibitem{piczak2015esc}
K.~J. Piczak,
\newblock ``{ESC}: Dataset for environmental sound classification,''
\newblock in {\em Proc. 23rd {ACM} Int. Conf. Multimedia ({MM})}, 2015, pp.
  1015--1018.

\bibitem{silero2024vad}
{Silero Team},
\newblock ``Silero {VAD}: Pre-trained enterprise-grade voice activity detector
  ({VAD}), number detector and language classifier,'' GitHub repository,
  \url{https://github.com/snakers4/silero-vad}, 2024.

\bibitem{lugosch2019fluent}
L.~Lugosch, M.~Ravanelli, P.~Ignoto, V.~S. Tomar, and Y.~Bengio,
\newblock ``Speech model pre-training for end-to-end spoken language
  understanding,''
\newblock in {\em Proc. Interspeech}, 2019, pp. 814--818.

\bibitem{busso2008iemocap}
C.~Busso et~al.,
\newblock ``{IEMOCAP}: Interactive emotional dyadic motion capture database,''
\newblock {\em Lang. Resour. Eval.}, vol. 42, no. 4, pp. 335--359, 2008.

\bibitem{nagrani2017voxceleb}
A.~Nagrani, J.~S. Chung, and A.~Zisserman,
\newblock ``{VoxCeleb}: {A} large-scale speaker identification dataset,''
\newblock in {\em Proc. Interspeech}, 2017, pp. 2616--2620.

\bibitem{panayotov2015librispeech}
V.~Panayotov, G.~Chen, D.~Povey, and S.~Khudanpur,
\newblock ``{Librispeech}: An {ASR} corpus based on public domain audio
  books,''
\newblock in {\em Proc. IEEE ICASSP}, 2015, pp. 5206--5210.

\end{thebibliography}
\end{document}